\documentstyle[12pt]{article}

\begin{document}
\baselineskip 0.5cm

\rightline{McGill/94-16}

\begin{center}
{\large\bf Probing quark gluon plasma with jets }
\vskip 1.0cm
{\bf Jicai Pan  \ and \ Charles Gale }
\vskip 0.6cm
{\it
Department of Physics, McGill University	\\
3600 University St., Montr\'eal, PQ, H3A 2T8 Canada
}
\end{center}

\vskip 0.6cm
\centerline{\bf Abstract}
\vskip 0.2cm

	We study multiple scatterings of jets on constituents
of quark gluon plasma and introduce
energy--energy correlations to quantify their effects.
The effects from a longitudinally expanding plasma on medium
as well as high energy
jets are found to be significant at both RHIC and LHC energies.
Because jets escape from the plasma long before the completion
of mixed phase, these effects are free from complications of
final state hadronic interactions and decays.
These  suggest that  jets can be used  to
probe the plasma that might be created in future high energy
heavy ion collisions.

PACS numbers: 25.75.+r, 12.38.Mh, 24.85.+p, 12.38.Bx

\newpage
\section{Introduction}

  In high energy nuclear collisions new states of matters,
e.g. the quark gluon plasma predicted by numerical calculations
of lattice gauge theories, might be produced. Because of the
many-body heavy-ion system and soft multihadronic
production, major efforts are to find experimental observables
that could provide information on the states of matters
produced in the collisions [1]. Many observables that have been
intensively studied are mostly products created in the plasma,
like dileptons, photons, $J/\psi$ and strange particles.
These processes can carry information about the plasma,
but some of them are usually complicated by final state
hadronic interactions and decays.

  In this paper we are interested in jets propagating
through the plasma. It is of great interest
for following reasons.
(a) Jets have been extensively investigated both
experimentally
and theoretically.
Because they are  produced by hard collisions of
quarks and gluons, they can be and have been
calculated with perturbative QCD to various
degrees of accuracy.
The agreement between theory and experiments shows
that jets are one of the
very few well understood high energy strong
interacting phenomena.
(b)Jets are produced at the very beginning
of the collision process
and they propagate in the plasma and interact strongly with
the plasma constituents before they escape. Because of these
properties they can be used to probe the plasma.
(c) Since the final state hadronic interactions
and decays take place
after the phase transition that is expected to
complete after the
jets escape from the plasma, the jet signals are
virtualy free of
 background of final state interactions and decays.
(d)Jets are easily accessible experimentally.
The typical jet energy is
greater than 10 GeV which is much larger than
the typical energies of
constituents in a thermalized plasma at the
temperature that can be
created in the future heavy ion collider,
such as the RHIC and the LHC.
Jets are therefore expected to escape from
the plasma as well identified objects.

	The energy--energy correlations
of hadrons produced in $e^+e^-$ collider were suggested
by Basham, Brown, Ellis and Love [2] to test
QCD, and were investigated by a number of
experimental groups [3]. In hadronic and
nuclear collisions jets are
produced by hard scattering of partons.
In a $2 \rightarrow 2$ process
the two jets are produced back-to-back
in their rest frame. Thus the QCD
contribution to the energy--energy correlations
is dominated by
$2 \rightarrow 3$ process. Ali, Pietarinen and
Stirling~[4] showed
that transverse energy--energy correlations
depend very weakly on the
structure function and the normalized correlation function is
approximately proportional to coupling constant $\alpha_s$.

	A jet produced at the beginning of a
collision may scatter many times
on plasma constituents before it escapes from the plasma.
The average transverse momentum gained in
the multiple scattering can be as high as a few GeV.
This effect on a
jet of energy of several tens GeV is of the order of
$\alpha_s$. In fact, as we shall see the multiple scattering
has significant effects on energy correlations and
dominates over the QCD contribution in some cases.

	The paper is organized as follows.
In section 2 we start with
definition of energy--energy correlation
and  derive a general formalism for evaluating the
scattering effects on  correlations. We then calculate these
effects in a longitudinal hydrodynamical model that describes
plasma expansion. Hadronization effects are
discussed at the end
of the section. In section 3 we present the numerical results
and discuss their dependence on colliding energy,
jet energy and
nuclear size. In section 4 we summarize our conclusions.

\section{Jets in quark gluon plasmas}

  In nuclear collisions at RHIC and LHC energies,
jet productions are
no longer rare events. High transverse energy jets stand high
beyond the associated soft productions and
can be readily identified experimentally. Low transverse
energy jets are, on the other hand,
burried under the
large amount of soft processes.
To reduce the soft background, one studies
high transverse energy jets. To see the effects of
plasma on high
$E_T$ jets, it is essential to find sensitive variables.
In the following,
transverse energy--energy correlations are introduced
to study such
effects. As we shall see that the plasma indeed has
significant effects on the correlations.

\subsection{Energy--energy correlations and effects of plasmas}

 As aforementioned, the energy--energy correlations
were introduced
in [2] to test the QCD in $e^+e^-$ collisions, and transverse
energy--energy correlations were suggested in [4]
to measure $\alpha_s$
in $p\overline{p}$ collision. In nuclear collisions
we can similarly
define the transverse energy--energy correlations for high $E_T$
jets as
\begin{eqnarray}
{1 \over \Sigma}{d\Sigma \over d\phi}(E_T) =
\langle {1\over \Delta \phi}
\sum_{i,j}^n { E_{Ti}E_{Tj}\over E_T^2}\rangle_{event}
\end{eqnarray}
Here $E_{Ti}$ is the transverse energy of particle i,
$\Sigma=\int_0^{2\pi} (d\Sigma/d\phi) d\phi$,
and $\Delta\phi$ is the
angle between the particle pair $(i,j)$ on the
transversal plane.
The $\langle \ldots\rangle_{event}$ means average
over many events. To
obtain higher statistics one might integrate over
a range of transverse
energies. In this case one has
\begin{eqnarray}
{1 \over \Sigma}{d\Sigma \over d\phi}(\Delta E_T) =
\int_{E_T^{min}}^{E_T^{max}}
 d E_T {d^2\Sigma\over dE_T d\phi} \, \,  /
\int_{E_T^{min}}^{E_T^{max}} d E_T {d\Sigma\over dE_T}
\end{eqnarray}

	The numerical calculation of correlation
function has been done in ref. [4]
based on the perturbative calculations of hard
$2\rightarrow 2$ and
$2\rightarrow 3$ processes in QCD [5].

	Now let us consider the effects of the plasma.
Because two-jet
production dominates over multiple-jet production we need only
to consider the plasma effects on two-jet events. Let
$p_\mu$ and $p'_\mu$ be the four momenta of a jet before and
after the multiple scattering off plasma constituents. In high
energy elastic scattering, the scattering cross section
depends only on
four--momentum transfer squared, or the transverse
momentum because
$-(p_\mu+p'_\mu)^2=4p^2\sin^2(\eta/2)\approx p_T^2$. The $p_T$
distribution of the jet decreases rapidly with increasing $p_T$
and vanishes in the backward hemisphere ($\eta >\pi/2)$.
We assume that $f_F(p_T)$ is the $p_T$ distribution
of one jet which
we define
as the forward jet and $f_B(p_T)$ is that of the other jet
in backward, and they satisfy the following
normalization condition
\begin{eqnarray}
\int {d^3{\bf p} \over E} f_F(p_T) =
	\int {d^3{\bf p} \over E} f_B(p_T) =1.	\label{d31}
\end{eqnarray}
Here $E=Q/2$ and $Q$ is the total jet energy.
We note
\begin{eqnarray}
\int {d^3{\bf p} \over E} f_F (p_T) = {1\over E}\int_0^E
   d^2{\bf p}_T f_F (p_T)\int_0^{\sqrt{E^2-p_T^2}} dp_z .
\end{eqnarray}
After integrating over $p_z$ one can expand
$\sqrt{E^2-p_T^2}$ as
$E(1-p_T^2/2E^2 +\cdots)$. As we shall see later
that the  effects  of the plasma on
jets are of the order of $\langle p_T \rangle/E$, we can neglect
$\langle p_T^2 \rangle /E^2$ and higher terms. That is
\begin{eqnarray}
\int d^2{\bf p}_T f_F(p_T)
 =\int d^2{\bf p}_T f_B(p_T) =1. \label{d32}
\end{eqnarray}

	Let $d\sigma/d\Omega_0$ be the  cross
section for two-jet production
in a nuclear collision, with $\Omega_0$ being the usual solid angle.
The double energy cross
section after the scattering
on plasma constituents can be written as
\begin{eqnarray}
{d^2\Sigma\over d\Omega_1 d\Omega_2}={1\over 2}\int
 	d\Omega_0{d\sigma\over d\Omega_0}
	\int  {p^2_1dp_1\over E_1}{p^2_2dp_2\over E_2}
	\left[ f({\bf p}_1;{\bf p}_0)f({\bf p}_2;-{\bf p}_0)+
	 f({\bf p}_2;{\bf p}_0)f({\bf p}_1;-{\bf p}_0) \right]
\end{eqnarray}
which can be rewritten as
\begin{eqnarray}
{d^2\Sigma\over d\Omega_1 d\Omega_2}={1\over 2}\int
 	d\Omega_0{d\sigma\over d\Omega_0}
	\int  {p^2_1dp_1\over E_1}{p^2_2dp_2\over E_2}
 \left[ f_F(p_{T1})f_B(p_{T2})+f_F(p_{T2})f_B(p_{T1}) \right]
\end{eqnarray}
After replacing $p$ with $p_{T}/\sin\eta$
($\eta$ is the scattering angle with respect to the original jet
axis)
in the forward hemisphere and
$p$ with $p_{T}/\sin (\pi-\eta)$ in the backward
hemisphere, we obtain
\begin{eqnarray}
{d^2\Sigma\over d\Omega_1 d\Omega_2}={1\over 2}\int
 	d\Omega_0{d\sigma\over d\Omega_0}
	\left[ 	F(\eta_1)F(\pi-\eta_2) +
		F(\pi-\eta_1)F(\eta_2)	\right]	\label{d33a}
\end{eqnarray}
where
\begin{eqnarray}
F(\eta) = {1\over E}\int  p^2dp f_F(p_T),
 \qquad\mbox{for $\eta <{\pi\over 2}$}	\label{d34}
\end{eqnarray}
  From (\ref{d31}) we see that
\begin{eqnarray}
\int  d\Omega F(\eta) =1.			\label{d35}
\end{eqnarray}
Using relation $p=p_T/\sin\eta$, (\ref{d34}) can be written as
\begin{eqnarray}
F(\eta) = {1\over E \sin^3\eta}\int_0^{E\sin\eta} p^2_Tdp_T
	 f_F(p_T).
\end{eqnarray}
When $\sin\eta>>\langle p_T\rangle/E$, we can use
(\ref{d32}) and obtain
\begin{eqnarray}
F(\eta) = {\langle p_T\rangle \over 2\pi E \sin^3\eta}  ,
		\qquad\mbox{for $\eta <{\pi\over 2}$} \label{d37}
\end{eqnarray}
Here $\langle p_T\rangle$ is the average transverse
momentum defined by
\begin{eqnarray}
\langle p_T\rangle = \int p_Td^2{\bf p}_T f(p_T). \label{d38}
\end{eqnarray}

	The integration in (\ref{d33a}) can be carried out
numerically~[6]. To obtain a simple analytical result, we
note that in nuclear collisions only high transverse energy jets
can be well identified. Also the study of
jet propagation in longitudinal
direction is very complicated by the fact that plasma expands in the
same direction. For these reasons we restrict
ourselves to the study of
transverse jets. Near the transverse direction
the $d\sigma/d\Omega_0$
varies slowly with $d\Omega_0$, while $F(\eta)$ varies rapidly near
$\eta=0$. Thus the dominating
contributions to the  integration in (\ref{d33a}) come from small
angular regions about the two detection directions $\Omega_1$
and $\Omega_2$.
As an example we consider the contribution from
integration over the cone about $\Omega_0\approx\Omega_1$
which implies $\eta_1\approx0$, $\eta_2\approx\phi$.
Here $\phi$ is the angle between $\Omega_1$ and $\Omega_2$
on the transverse plane. Since the $F(\eta_1)$
changes rapidly, we can expand $d\sigma/d\Omega_0$
in the integrand
as a Taylor series in $\eta_1$ for fixed $\phi$. The dominating
contribution to the integral over this region is then obtained by
taking $d\Omega_0\approx d\Omega_1$,
${d\sigma / d\Omega_0}\approx {d\sigma / \Omega_1}$
and $\eta_2\approx\phi$. The integral over
$\Delta \Omega_1$ becomes
\begin{eqnarray}
{d\sigma\over d\Omega_1}F(\pi-\phi)
	\int_{\Delta \Omega_1} d\Omega_1 F(\eta_1)
	={d\sigma\over d\Omega_1}F(\pi-\phi),
		\qquad\mbox{for $\phi >{\pi\over 2}$}
\end{eqnarray}
Similarly, the integration over other cones
can be evaluated. Summing over all contributions
we obtain the double energy cross section
\begin{eqnarray}
{d^2\Sigma\over d\Omega_1 d\Omega_2}={1\over 2}
	F(\pi-\phi)
	\left[ {d\sigma\over d\Omega_1}+
	{d\sigma\over d\Omega_2} \right],
			\qquad\mbox{for $\phi >{\pi\over 2}$}
\end{eqnarray}
Integrating over double solid angle and keeping $\phi$
constant we obtain the energy--energy correlation
\begin{eqnarray}
{1\over \Sigma} {d\Sigma\over d\phi}=
	{2 \langle p_T\rangle \over \pi Q}
	{1\over \sin^3(\pi-\phi)},
	\qquad\mbox{for $\phi >{\pi\over 2}$}, \label{c31}
\end{eqnarray}
and there is no correlation for $\phi< \pi/2$ in
the leading order
approximation. Here $Q$ is the total jet energy
on the transverse plane.
The asymmetry about $\phi=\pi/2$ comes from the
following fact. Because the
$f_F(p_T)$ and $f_B(p_T)$ decrease rapidly
with increase scattering angles with respect
to two initial back-to-back jets,
the correlation is the strongest at $\phi=\pi$
and decreases monotonously
with decreasing $\phi$ and eventually vanishes at $\phi=0$.
The discontinuity of about $\phi=\pi/2$
occurs because of the approximation.
To avoid the discontinuity we can extrapolate the
correlation to the region $\phi<\pi/2$ with
\begin{eqnarray}
{1\over \Sigma} {d\Sigma \over d\phi}=
	{2 \langle p_T\rangle \over \pi Q}\sin\phi ,
 \qquad\mbox{for $\phi \leq {\pi\over 2}$}. \label{c31a}
\end{eqnarray}
This  is sensible because in $\phi<\pi/2$ the correlation
decreases monotonously to zero at $\phi=0$ as it should be,
and the correlation function
as well as its first derivative are continuous at $\phi=\pi/2$.

\subsection{Multiple scattering in expanding plasmas}

	Now we turn to the calculation of $\langle p_T\rangle$.
Let $g({\bf k}_T)$ be the normalized
transverse momentum distribution of
a jet after {\it one} \ scattering with a plasma constituent
and $P(n)$ be the normalized probability that a
jet scattering $n$ times with the constituents.
The transverse momentum distribution of a jet after multiple
scattering with plasma constituents is then
\begin{eqnarray}
f(p_T)=\sum_{n=0} P(n)\int\prod_{i=1}^{n}d^2{\bf k}_{Ti}
	g({\bf k}_{T1})\cdots g({\bf k}_{Tn})
	\delta({\bf p}_{T}-\sum_{j=1}^n{\bf k}_{Tj}),
\end{eqnarray}
with the understanding that
$f({\bf p}_T)=\delta({\bf p}_T)$ when $n=0$.
In an equilibrium plasma the multiple scatterings are independent
from one another. In this case one can easily show
\begin{eqnarray}
\langle p_{T}^2 \rangle =N \langle k_{T}^2 \rangle,
\end{eqnarray}
where $\langle p_T^2 \rangle$ is the average transverse momentum
after {\it multiple} scatterings  defined as
\begin{eqnarray}
\langle p_T^2 \rangle=\int d^2{\bf p}_T p^2_T f_F({\bf p}_T),
\end{eqnarray}
and $\langle k_T^2 \rangle$ is the average transverse momentum
after {\it one} scattering defined as
\begin{eqnarray}
\langle k_T^2 \rangle=\int d^2{\bf k}_T k^2_T g({\bf k}_T),
\end{eqnarray}
and $N$ is the average number of scatterings given by
$N=\sum_{n=0} n P(n)$.
Noting that the $\langle p_T^2 \rangle$ is
proportional to
$\langle p_T \rangle^2$, and $\langle k_T^2 \rangle$
to $\langle k_T \rangle^2$, and both distribution should
be similar in shape, we expect
\begin{eqnarray}
\langle p_{T} \rangle =\sqrt{N} \langle k_{T} \rangle,\label{q31}
\end{eqnarray}

In hydrodynamical models $N$ can be calculated from
\begin{eqnarray}
N=\int dx \left[ \sum_i n_i(x)\sigma_i \right].	\label{q32}
\end{eqnarray}
Here $n_i$ is the particle density of the ith plasma constituent,
$\sigma_i$ is the scattering cross section of a
jet with the ith constituent. The integral is carried out
along the jet trajectory.

	We consider central  nuclear collisions where
the created plasma is constrained in a cylinder of
radius $R=1.2A^{1/3}$ (fm) where $A$
is the nuclear mass number. The evolution of the plasma
after its formation is
described by hydrodynamical equations.
In the Bjorken model [7] the system
expands isentropically along the longitudinal
direction and
the entropy density decreases according to
\begin{eqnarray}
s(\tau)=s(\tau_i)\tau_i/\tau		\label{q33}
\end{eqnarray}
while the temperature falls according to
\begin{eqnarray}
T(\tau)=T_i\left[\tau_i/\tau\right]^{1/3}	\label{q34}
\end{eqnarray}
where $\tau_i$ is the proper time at which the plasma is
formed and $T_i$ is
its initial temperature. When the temperature drops
to $T_c$ at time $\tau_c$
a mixed phase emerges. In the end of the mixed phase
at time $\tau_h$
the system is fully converted into hadrons. It can
be shown that
\begin{eqnarray}
\tau_h=r\tau_c					\label{q35}
\end{eqnarray}
where $r$ is the ratio of degree of freedoms in the
plasma phase to that in the hadronic phase.

	Now let us rewrite (\ref{q32}) as
\begin{eqnarray}
N=\sigma_{pl}\int d\tau n(\tau)
\end{eqnarray}
where $\sigma_{pl}$ is the averaged cross section and $n(\tau)$ is the
total particle density in plasma.
In the additive quark model the quark--quark
crosss section is $\sigma_{qq}\approx 4$ mb and in perturbative
QCD one has
$\sigma_{gg}=(9/4)\sigma_{qg}=(9/4)^2\sigma_{qq}$.
Hence, we might expect
$\sigma_{pl}\approx (16+4.6N_f)/(16+12N_f)\sigma_{gg}\approx 12 $ mb.
Noting that for an idea gas the particle density
is proportional to the entropy
density and using the expansion equation (\ref{q33}) we have
\begin{eqnarray}
N=\sigma_{pl}\tau_i n(\tau_i)\ln(\tau_f/\tau_i)	\label{q36}
\end{eqnarray}
where $n(\tau_i)$ is the initial particle density.
$\tau_f$ is the time
for the jets to escape from the plasma,
if this happens before beginning of the mixed phase,
i.e. $\tau_f<\tau_c$.
When $\tau_f>\tau_c$ jets will interact with
the mixed phase constituents:
quarks, gluons and hadrons.

	Because of the large degree of freedom
in the plasma phase (c.f. eq.~\ref{q35}),
the mixed phase is long
and jets propagating transversal
to collision direction are expected to escape
from the plasma before the completion of mixed phase.
Thus hadronic gas has no effects on jets.

	We note $\sigma_{q\pi}=(2/9)\sigma_{pp}\approx 9$ mb.
For simplicity  we assume
$\sigma_{q\pi}=\sigma_{pl}$.
In this case the number of collisions can still
be expressed by (\ref{q36}). There $\tau_f$
is the escape time of jets which
may have been scattered by mixed phase constituents.
When a jet pair is created
at a distance $R_0$ from the collision axis, the
average number of collisions
each jet experienced is
\begin{eqnarray}
N={1\over 2}\sigma_{pl}\tau_i n(\tau_i)\ln{R^2-R_0^2\over \tau_i^2}
\end{eqnarray}
and $N=0$ if $R^2-R_0^2\leq \tau_i^2$.
Since $R_0$ is unmeasurable, an average over geometry must be made.
For a sharp nuclear density profile we obtain
\begin{eqnarray}
N={1\over 2}\sigma_{pl}\tau_i n(\tau_i)
	\left[ \ln{R^2\over \tau_i^2} -{1\over 2}+
	{\tau_i^4\over 2 R^4}\right]
\end{eqnarray}
Noting that $n(\tau_i)=s(\tau_i)/c$ with $c=3.6$, we obtain
\begin{eqnarray}
N={g_Q\pi^2\over 45 c}\sigma_{pl}\tau_iT^3_i
	\left[ \ln{R^2\over \tau_i^2} -{1\over 2}+
	{\tau_i^4\over 2 R^4}\right]			\label{q37}
\end{eqnarray}
where $g_Q=16+21N_f/2$ is the degree of freedom in plasma and $N_f$ is
the number of active quark flavor.

  In the above calculations only longitudinal
expansion is taken into
account. When the system expands transversally
as well, it will take
longer for jets to escape from the plasma.
This may increase the number of
collision and enhance plasma effects. On the other hand, transverse
expansion accelerates the cooling process and reduces plasma density
more rapidly. Hence, the resulting effects is expected to be small.
Furthermore, we note that transverse expansion
is negligible in the early stage of the expansion.
As a result, the transverse expansion has little effects on jets.

\subsection{Hadronization}

	The results obtained so far are for quarks and gluons.
To compare with experiments, the effects of hadronization
should be taken into account.
Basham et al. [2] have calculated that the hadronization
correction to  the energy pattern and correlations.
Similarly one can show that the hadronization correction
to  the transverse energy--energy correlations is given by
\begin{eqnarray}
{ c_0 \langle h_T\rangle \over \pi Q\sin^3\phi} \ ,
\qquad\mbox{for $\phi <\pi$}.			\label{h2}
\end{eqnarray}
Here $c_0=2.5$ is the mutiplicity density at rapidity $y=0$ and
$\langle h_T\rangle$ is  the average
transverse momentum of hadrons from jet fragmentation [2].
Because two hadrons with angle $\phi$ may come from either the same
jet or from the two separate back-to-back jets,  the hadronization
correction to correlation is symmetric about $\phi=\pi/2$.

	Summing over the effects of QCD corrections, multiple
scattering in plasma and hadronization, we finally obtain the
energy--energy correlation
\begin{eqnarray}
{1 \over \Sigma}{d\Sigma\over d\phi}=
	{1 \over \Sigma}{d\Sigma^{(QCD)}\over d\phi}   +
	\left\{
	\begin{array}{ll}
	{\displaystyle
	{2 \sqrt{N}\langle k_T\rangle +c_0 \langle h_T\rangle
	 \over \pi Q\sin^3(\pi-\phi)}
	} 	& \mbox{for $\phi > \pi/2$}	\\ 	\\
	{\displaystyle
	{2\sqrt{N} \langle k_T\rangle \sin\phi \over \pi Q}  +
	{c_0 \langle h_T\rangle \over \pi Q \sin^3\phi}
	}	& \mbox{for $\phi \leq  \pi/2$}.
	\end{array}
		\right.				\label{res}
\end{eqnarray}

\section{Results and discussion}
	From (\ref{res}) and (\ref{q37})
we see that the results depend on three parameters, the
initial temperature $T_i$, the jet energy $Q$ and the
geometrical size $R$, that can be changed in experiments.

	We first look at central Au--Au collisions at RHIC
energy (200 GeV
per nucleon) where the initial temperature is expected to be
about 250 MeV. The results for correlations are shown in
Figs.~1 at jet energies $Q=10$, and 20 GeV. In the calculation we
used $N_f=3$, $\sigma_{pl}=12$ mb, $\Lambda=0.2$ GeV, $\tau_i=1$ fm,
$\langle k_T\rangle = \langle h_T\rangle=0.36$ GeV and the results of
refs. [4, 5] for $d\Sigma^{(QCD)} / d\phi$.
As we can see from the figure
that the hadronization effects are unimportant at $Q=20$ GeV, while
the plasma effects are distinctly beyond the QCD effects.
To minimized the hadronization
effects one studies the asymmetric correlations defined as [2]
\begin{eqnarray}
A(\phi)={1 \over \Sigma}\left[ 	{d\Sigma \over d\phi}(\pi-\phi) -
 {d\Sigma \over d\phi}(\phi) \right] \ ,
 \qquad\mbox{for $\phi < \pi/2$}
\end{eqnarray}
In Fig.~2 we show the corresponding $A(\phi)$ distributions.
 From the figure we see that the  plasma effects are
significant and  should be easily measured experimentally
if the plasma is created in the collision. Here the hadronization
effects cancel  because of the symmetry about
$\phi =\pi/2$.

	In Figs.~3 and 4 we show the
$(1/\Sigma)(d\Sigma/d\phi)$ and
$A(\phi)$, respectively, for $Q=40$,and 100 GeV at LHC energy
(6300 GeV per nucleon)
where we estimate  $T_i=500$ MeV.
Compare Fig. 1 with 3 and 2 with 4 we
see that plasma effects increase with increasing initial
temperature. At LHC the plasma effects stand well beyond
the effects of QCD corrections and hadronizations for
jet energy as high as 100 GeV.

	To see the plasma effects more directly,
one might compare the
integrations that are defined as
\begin{eqnarray}
\sigma_C=\int_{\phi_1}^{\phi_2} {1 \over \Sigma}
		{d\Sigma \over d\phi}(\phi)  d \phi
\end{eqnarray}
for energy--energy correlations and,
\begin{eqnarray}
\sigma_A=\int_{\phi_1}^{\phi_2}A(\phi) d \phi
\end{eqnarray}
for asymmetric energy--energy correlations. The results for various
$T_i$ and $Q$ are listed in Tables~1 and 2 respectively.
Here we integrated over $\phi$ from $\pi/6$ to
$5\pi/6$ for $\sigma_C$,
and from $\pi/6$ to $\pi/2$ for $\sigma_A$.
{}From the tables we see that
the integrations are enhanced by a factor
of 1.2 to 1.6 due
to the presence of the plasma. In $e^+e^-$ experiments
the integration
can be very well determined [8]. Similarly,
we expect that such enhancement
can be unambiguously measured in high $Q$ jet events
in heavy ion collisions.

\vskip 0.4cm
\begin{table}
\caption{Integrations of energy--energy correlation
	function for various $T_i$ and Q}
\vskip 0.4cm
\begin{tabular}{|c|c|c|c|c|} 				\hline
$T_i$ (MeV) & $Q$ (GeV) & QCD & QCD+Hadr. & QCD+Hadr.+Plasma \\ \hline
250	    &10	& 1.29	&	1.54	&  2.27		\\ \hline
250	    &20	& 1.32	&	1.44	&  1.80		\\ \hline
500	    &40	& 0.93	&	0.99	&  1.51		\\ \hline
500	    &100& 0.95	&	0.98	&  1.19		\\  \hline
\end{tabular}
\end{table}

\vskip 0.4cm
\begin{table}
\caption{Integrations of asymmetric energy--energy
	correlation function for various $T_i$ and Q}
\vskip 0.4cm
\begin{tabular}{|c|c|c|c|c|} 			\hline
$T_i$ (MeV) &$Q$ (GeV) & QCD 	& QCD+Hadr.  & QCD+Hadr.+Plasma \\ \hline
250	    &10	& 0.76	&	0.76	&   1.12	\\ \hline
250	    &20	& 0.33	&	0.33	&   0.52	\\ \hline
500	    &40	& 0.54	&	0.54	&   0.80	\\ \hline
500	    &100& 0.24	&	0.24	&   0.35	\\  \hline
\end{tabular}
\end{table}
\vskip 0.4cm

	In central collisions the initial particle
density increases with
$A^{1/3}$. When $R>>\tau_i$ the number of scatterings increases with
$A^{1/3}\ln (R^2/\tau^2_i)$. In Figs. 5 and 6 the
$(1/\Sigma)(d\Sigma/d\phi)$ and $A(\phi)$ distributions for $Q=40$ GeV
for S--S collision are compared with those for Au--Au collision at LHC,
respectively.
We see that the results depend rather weekly on the nuclear geometry.

	We note that in the kinematic region that we are interested in,
dominating effects are from scatterings of two gluon jets, and our results
are not applicable to scattering angle
$\sin\eta \/ _{\stackrel{<}{\sim}}\langle p_T\rangle/Q$
which corresponds to the region of $\phi\approx 0$ and $\phi \approx \pi$.

	In small scattering angle the multiple soft gluon
bremsstrahlung
plays an important role [9,10].
In the regions of $\phi\approx 0$ and $\phi \approx \pi$ one might
study the momentum imbalance or acoplanarity as suggested by Appel [9],
Blaizot and McLerran [10].

\section{Conclusions}
	We used (asymmetric) transverse energy--energy correlations
to study the
effects of multiple scatterings of jets on constituents of
quark gluon plasma.
In the calculations we considered the expansion of plasma,
QCD corrections
and nonperturbative effects. Our results show that the scattering effects
are remarkably strong in (asymmetric) correlations.
At LHC they stand well
above the background from  QCD corrections and hadronizations
for jet energy
as high as 100 GeV. If the plasma is formed in nuclear collisions and go
through a phase transition, the jets are expected to escape before the
completion of mixed phase due to the large degree of freedom in plasma.
In this case the observed effects are free of final
state hadronic interactions
and decays. These imply that jets could be used to
probe the quark gluon plasma.

	The plasma effects increases with decreasing jet energy.
To obtain maximal effects, jet energy should be kept as low as possible.
We note, however, that low energy jets, or minijets, can be
created or quenched
due to multiple interactions with plasma constituents.
Because of fluctuation
in soft processes [11], it is also elusive to identify
minijets in experiments.
To avoid such backgrounds and experimental ambiguities,
jet energy should be kept
high enough. Total jet energies of $Q=10$ and 40 GeV might be s
ensible thresholds
at RHIC and LHC energies, respectively.

  The main background of jet signature is probably the hadronic gas.
When jets propagate in hadronic gas, the multiple
scattering between jets and hadrons results
in similar effects as plasmas do. In general
one expects that hadronic gas exists
at lower temperature. In very high energy heavy ion collisions,
such as RHIC and LHC, the effects from hadronic gas should be
smaller than those from plasmas. To make quantitative comparison,
realistic models and detailed calculations are necessary~[6].

\section*{Acknowledgments}
We woulk like to thank H.C. Eggers for useful discussions.
This work was supported in part by the National Sciences
and Engineering Research
Council of Canada, and in part by the FCAR fund of the
Qu\'ebec Government.

\newpage
\section*{Figure captions}
\begin{description}
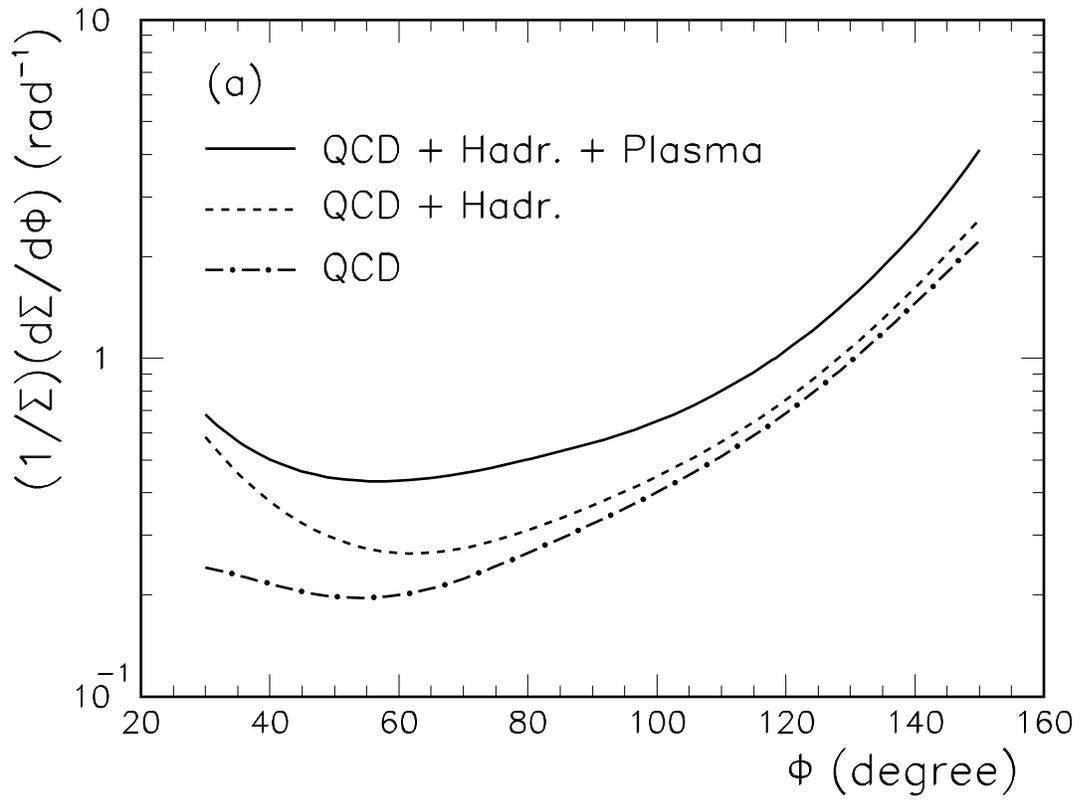
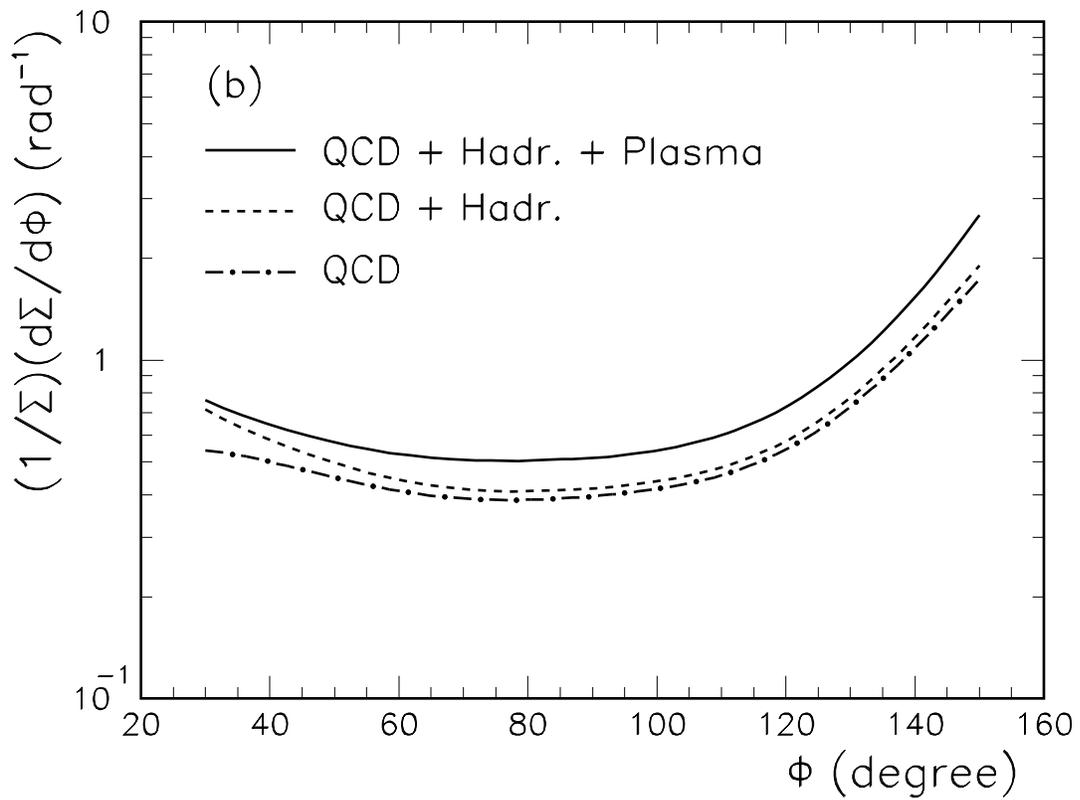
\item[Fig. 1] The $(1/\Sigma)(d\Sigma/d\phi)$ distribution
	for $T_i= 250$ MeV
	in Au--Au collisions. (a) for total transverse
	jet energy $Q=10$ GeV, and (b) for $Q=20$ GeV.

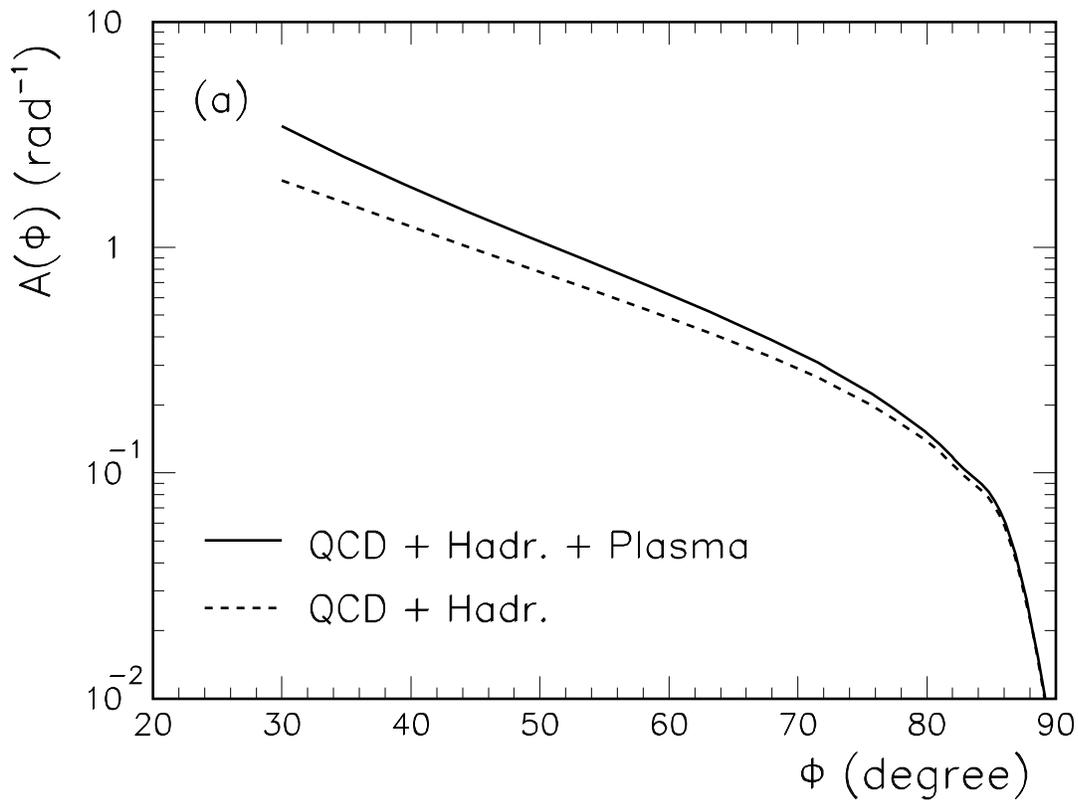
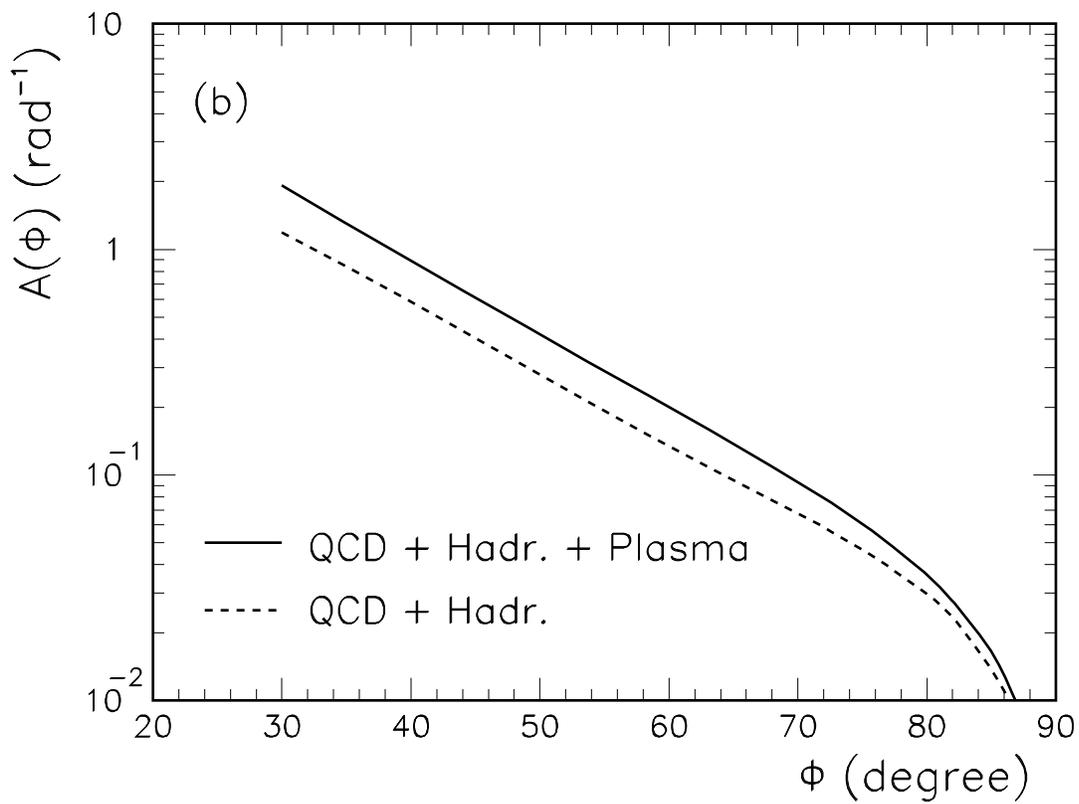
\item[Fig. 2] The $A(\phi)$ distribution for the same
	parameters as those in
	Fig.~1. (a) for $Q=10$ GeV, and (b) for $Q=20$ GeV.

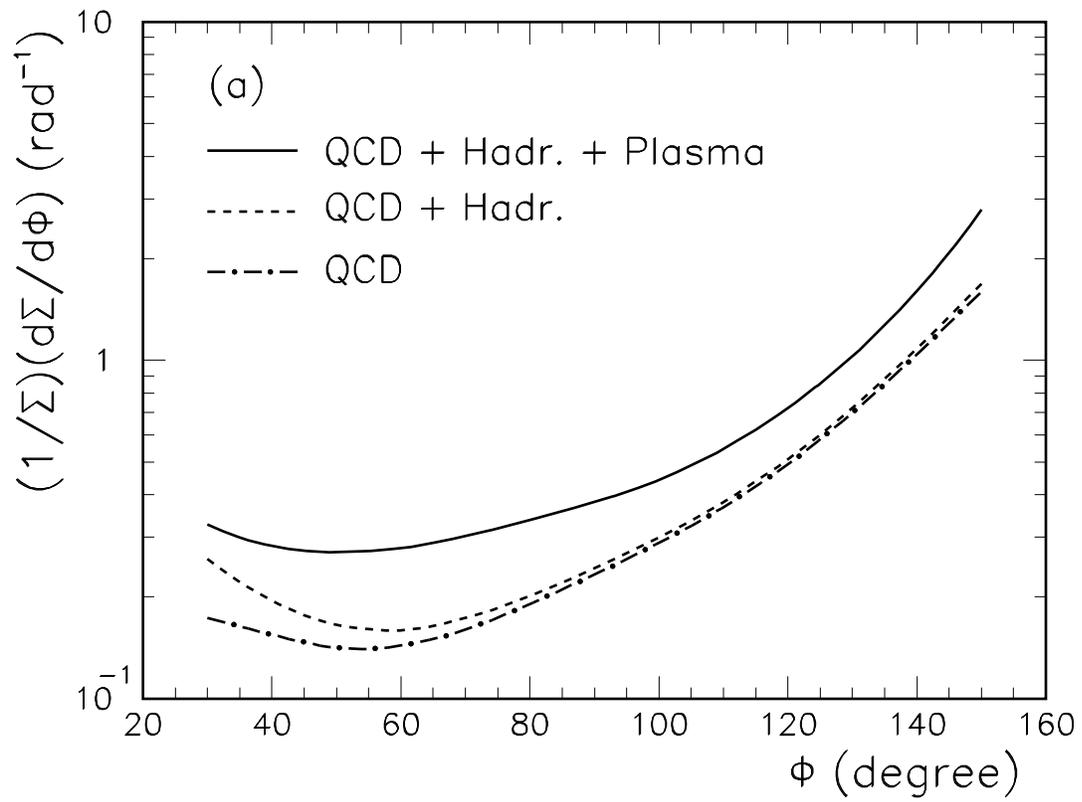
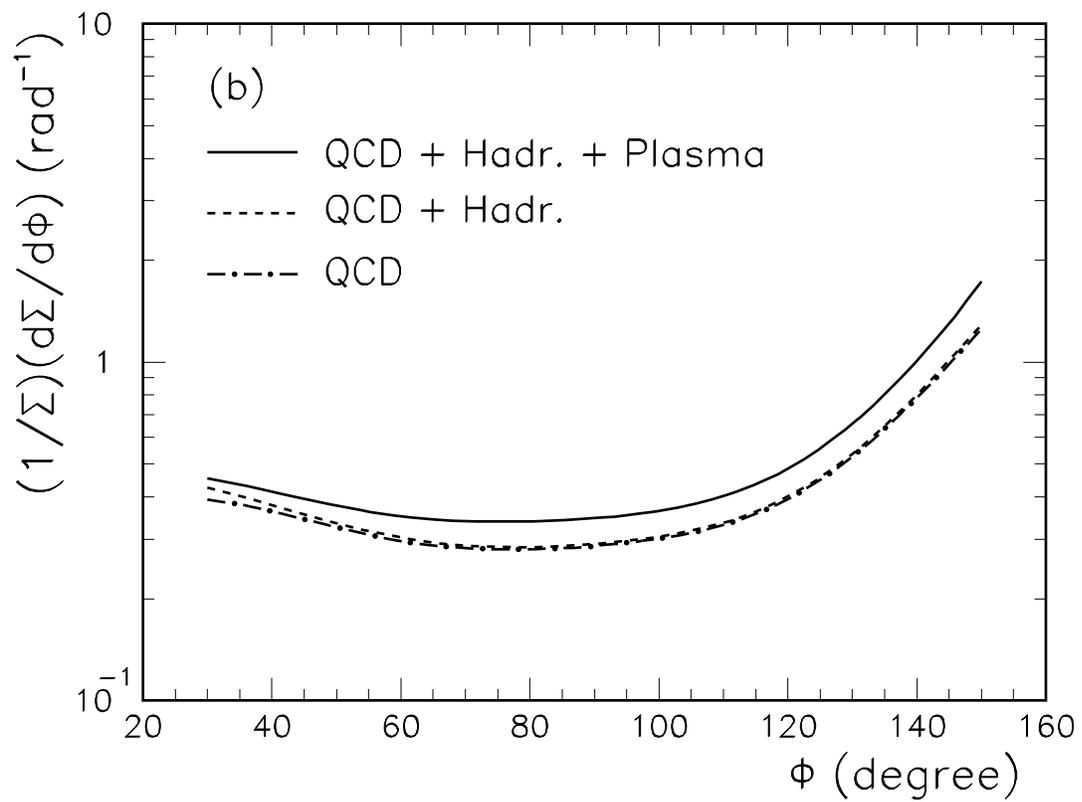
\item[Fig. 3] The $(1/\Sigma)(d\Sigma/d\phi)$ distribution
	for $T_i= 500$ MeV
	in Au--Au collisions. (a) for $Q=40$ GeV,
	and (b) for $Q=100$ GeV.

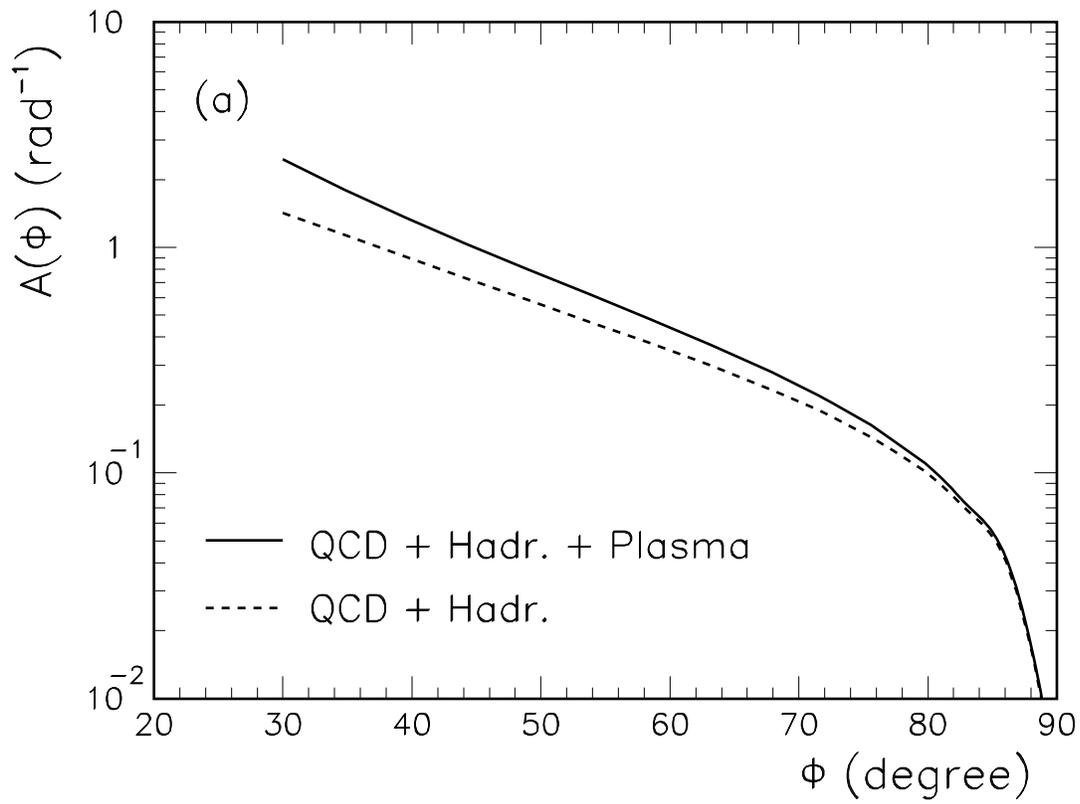
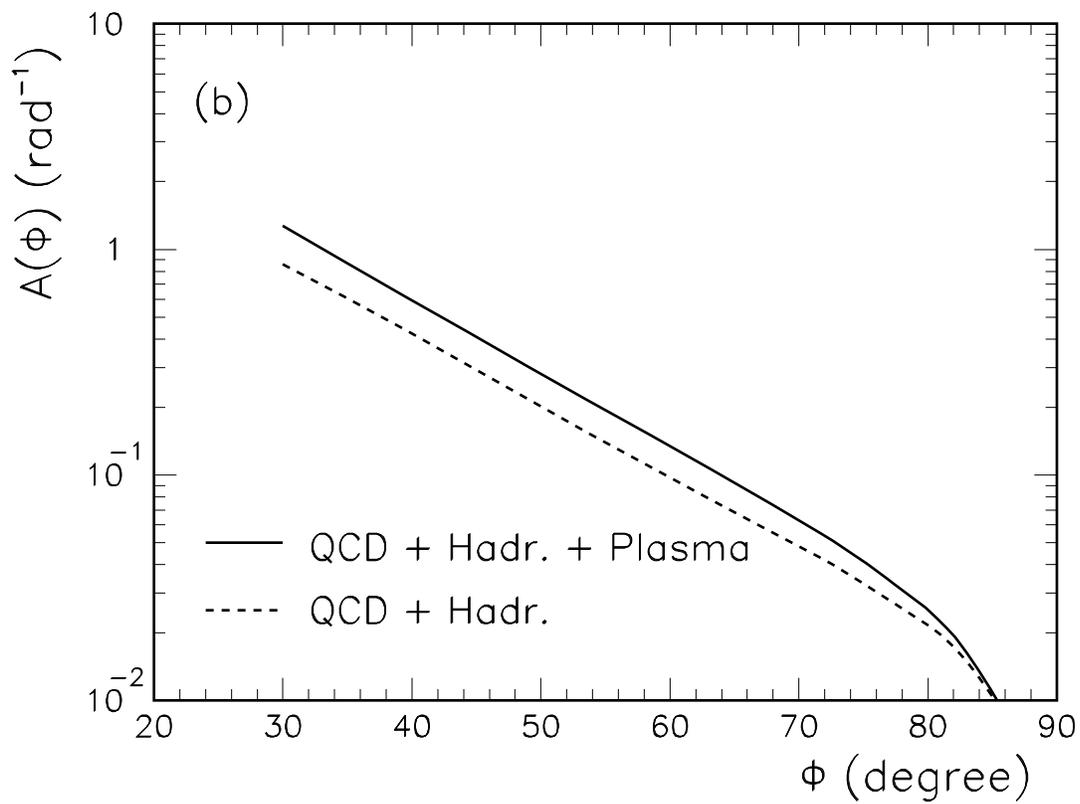
\item[Fig. 4] The $A(\phi)$ distribution for the same
	parameters as those in
	in Fig.~3. (a) for $Q=20$ GeV, and (b) for $Q=100$ GeV.

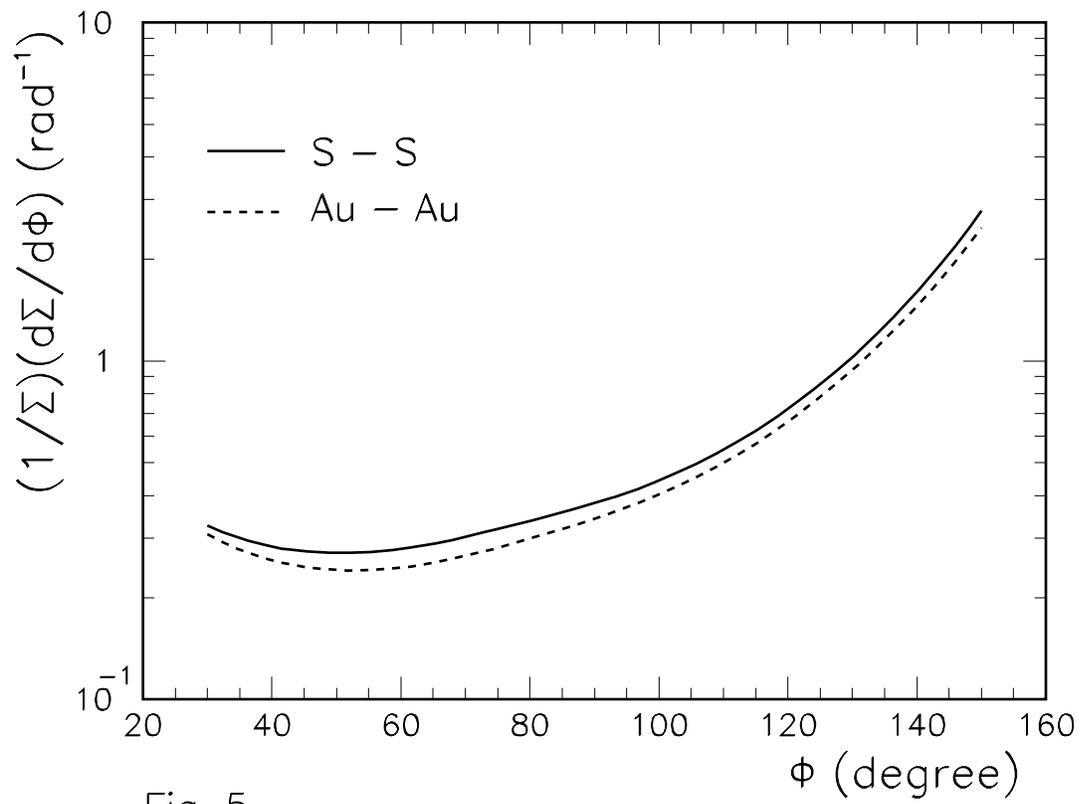
\item[Fig. 5] The comparison of $(1/\Sigma)(d\Sigma/d\phi)$
	distribution for
	S--S and Au--Au collisions for $Q=40$ GeV
	at $T_i=500$ MeV in Au--Au collisions.

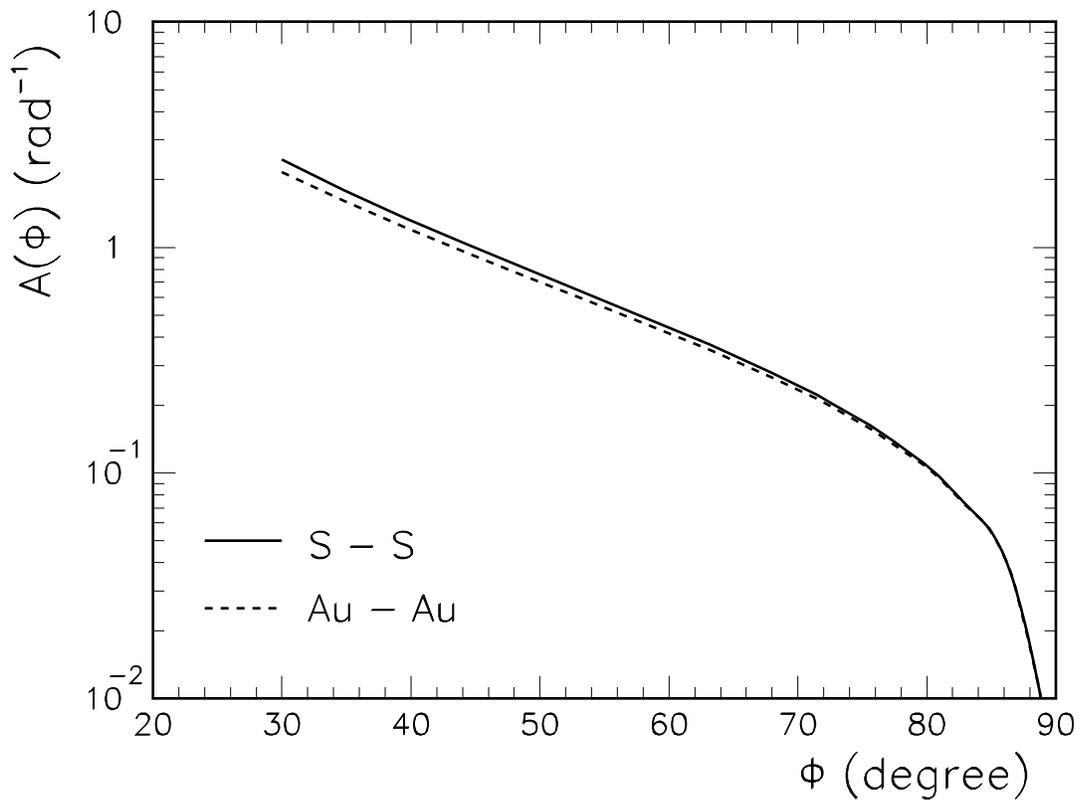
\item[Fig. 6] The comparison of $A(\phi)$ distribution
	for S--S and Au--Au
	collisions for the same parameters as those in Fig. 5.
\end{description}


\newpage
\begin{thebibliography} {99}
\bibitem{qm} See, e.g., Advanced Series on Directions in High
	Energy Physics, {\it Quark Gluon Plasma},
	edited by R. C. Hwa, (World Scientific, 1990);
	proceedings of {\it Quark Matter '91},
	Gatlinburg, Tennessee, USA, 1991, edited by
	T.C. Awes, at al. (North--Holland, 1992);
	proceedings of {\it Quark Matter '93},
	Borl\"ange, Sweden, 1993, (to be published);
	and references therein.
\bibitem{basham} C.L. Basham, L.S. Brown, S.D. Ellis and S.T. Love,
	Phys. Rev. D{\bf 17}, 2298 (1978),
	Phys. Rev. Lett. {\bf 41}, 1585 (1978),
	Phys. Rev. D{\bf 19}, 2018 (1979),
	Phys. Rev. D{\bf 24}, 2382 (1981).
\bibitem{abreu} See for examples:
	P.  Abreu et al. (DELPHI Coll.),
	Phys. Lett. B{\bf 252 }, 149 (1990),
	M.Z.Akrawy et al. (OPAL Coll.),
	Phys. Lett. B{\bf 252 }, 159 (1990),
	B.Adeva et al. (L3 Coll.),
	Phys. Lett. B{\bf 257}, 469 (1991), and
	references cited therein.
\bibitem{ali} A. Ali, E. Pietarinen and W.J. Stirling,
	Phys. Lett. B{\bf 141}, 447 (1984).
\bibitem{kirp} J. Kripfganz and A. Schiller,
	Phys. Lett. B{\bf 79}, 317 (1978);
	A. Schiller, J. Phys. G5, 1329 (1979);
	T. Gottschalk and D. Sivers,
	Phys. Rev. D{\bf 21}, 102 (1980);
	F.A. Berends et al. Phys. Lett. B{\bf 103}, 124 (1981);
	Z. Kunszt and E. Pietarinen,
	Nucl. Phys. B{\bf 164}, 45 (1980);
	Phys. Lett. B{\bf 132}, 453 (1983).
\bibitem{pg}  Jicai Pan and Charles Gale, in preparation.
\bibitem{bj}  J.D. Bjorken, Phys. Rev. D{\bf 27}, 140 (1983).
\bibitem{appel}  D. Schlatter et al.,
	Phys. Rev. Lett {\bf 49}, 521 (1982), and [2,3]
\bibitem{appel}  D.A. Appel, Phys. Rev. D{\bf 33}, 717 (1986).
\bibitem{blaizot}  J.P. Blaizot and L. McLerran ,
	Phys. Rev. D{\bf 34}, 2739  (1986).
\bibitem{bj} W.Q. Chao, T. Meng and J. Pan ,
	Phys. Rev. Lett. {\bf 58}, 1399 (1987);
 	T. Meng and J. Pan , Phys. Rev. D{\bf 37}, 243 (1988).

\end{thebibliography}
\end{document}